\documentclass{article}

 \usepackage[dblblindworkshop, final]{neurips_2026}
\workshoptitle{MLforSystems and AI4Good}

\usepackage[utf8]{inputenc} 
\usepackage[T1]{fontenc}    
\usepackage{hyperref}       
\usepackage{url}            
\usepackage{booktabs}       
\usepackage{amsfonts}       
\usepackage{nicefrac}       
\usepackage{microtype}      
\usepackage{xcolor}         

\usepackage{adjustbox}
\usepackage{comment}
\usepackage{wrapfig}
\usepackage{multirow}
\usepackage{titlesec}

\newcommand{\topic}[1]{\noindent\textbf{#1}}

\titlespacing*{\section}
  {0em}{0.3em}{0.3em}

\titlespacing*{\subsection}
  {0em}{0.2em}{0.2em}

\title{
    How Do Hardware Faults Manifest Differently \\in LLM Agents?}

\author{%
    Aiden Rubey\thanks{%
        Both authors are contributed equally to this work.},
    Zachary Coalson\footnotemark[1],
    $^{\dagger}$Cong Chen,
    $^{\dagger}$Jean-Philippe Monteuuis, \\
    $^{\dagger}$\textbf{Jonathan Petit,
    and Sanghyun Hong} \\
    Oregon State University, $^{\dagger}$Qualcomm Inc.
}

\begin{document}

\maketitle

\begin{abstract}
We present the first systematic study of
the resilience of large language model (LLM) agents
to random hardware-induced faults.
Prior work on hardware fault resilience in LLMs
focuses on model-level outcomes, such as
incorrect outputs or generation failures.
Yet LLM agents repeatedly interact with tools and environments,
creating loops that can change how hardware faults manifest.
We conduct an extensive evaluation across three agent benchmarks,
comprising 6,814 agent executions:
2,400+ clean runs and 4,300+ fault-injected runs 
with independently sampled random two-bit weight faults.
We trace how these faults propagate through agent execution.
Unlike prior studies,
we find that agentic execution transforms
model-level corruption into distinct system-level outcomes:
(1) agents can recover from faulty model outputs
and still complete tasks correctly;
(2) this recovery increases computational and execution costs;
(3) a few faults cause subtle but semantically meaningful errors 
while the agent otherwise behaves normally; and
(4) for some models and scenarios,
faults can alter security and alignment behavior.
\end{abstract}

\section{Introduction}
\label{sec:intro}

Large language models (LLMs) are increasingly deployed 
not only as standalone inference services,
but also as the decision-making components of autonomous agents.
These agents have enabled ambitious applications,
from autonomous science~\citep{gottweis2026coscientist, lu2026towards}         
and software engineering~\citep{yang2024sweagent, anthropic2025claudecode, openai2025codex}        
to web automation~\citep{zhou2023webarena, he2024webvoyager, skyvern2024},              
by repeatedly invoking LLMs to reason, interact with tools,
observe the results, and decide what to do next.
Serving such agents at scale relies on accelerator-rich platforms
with large, heterogeneous memory hierarchies,
where hardware faults remain an important reliability concern~\citep{xie2026strata, lei2026aegis, zheng2026sdcs}.

Research has shown that random hardware faults
can cause generation failures, degrade output quality,
or introduce subtle errors in otherwise plausible LLM outputs%
~\citep{agarwal2023resilience, sun2025demystifying, huang2026not}.
%
While these studies have expanded our understanding
of fault propagation within LLM inference,
they largely stop at the model output: once a corrupted generation is produced, 
it is treated as the final (and only) outcome of the fault.
It therefore remains unclear how the same corruption propagates
when that output becomes an intermediate step in a longer agent execution.

In this work, we present the first systematic study of 
the resilience of LLM agents to random hardware-induced faults.
Specifically we ask: 
\emph{what happens when a model output corrupted by a random hardware fault
enters the execution trajectory of an agent?}
Agent execution forms a \emph{closed loop}:
a corrupted output can trigger an action,
whose result becomes context for subsequent steps.
This extends fault propagation beyond the model output
into the agent's execution and environment state.

To answer this question,
we conduct an extensive evaluation across three agent settings.
We inject randomly sampled two-bit faults into LLM weights 
and measure their impacts over complete agent executions.
We use two-bit corruption as a controlled fault model,
following prior resilience studies~\citep{sun2025demystifying, huang2026not} that model memory errors beyond the correction capability of single-error-correcting ECC deployed in modern GPUs~\citep{oles2024understanding, zhu2025understanding, nvidiaECC}.
We focus on model weights, as they are reused 
across successive LLM invocations, 
allowing us to study how corruption in persistent model state 
propagates through closed-loop agent execution.

Our results show that model-level corruption and agent-level impact can differ.
(1) Agents can recover from corrupted intermediate outputs 
and still complete the task correctly, 
showing that a faulty output does not necessarily become a task failure. 
(2) This recovery comes at a cost: 
faulty executions can increase 
mean output tokens per execution by 38.0\%. 
(3) Not all faults are obvious.
These faults produce coherent executions with subtly incorrect action semantics,
such as wrong numerical values or action parameters.
(4) The impact of faults can extend beyond task correctness: 
for particular models and scenarios, 
we observe changes in security- and alignment-relevant behavior.

\section{Background and Related Work}
\label{sec:prelim}

\begin{figure}[t]
\centering
\includegraphics[width=1.0\linewidth]{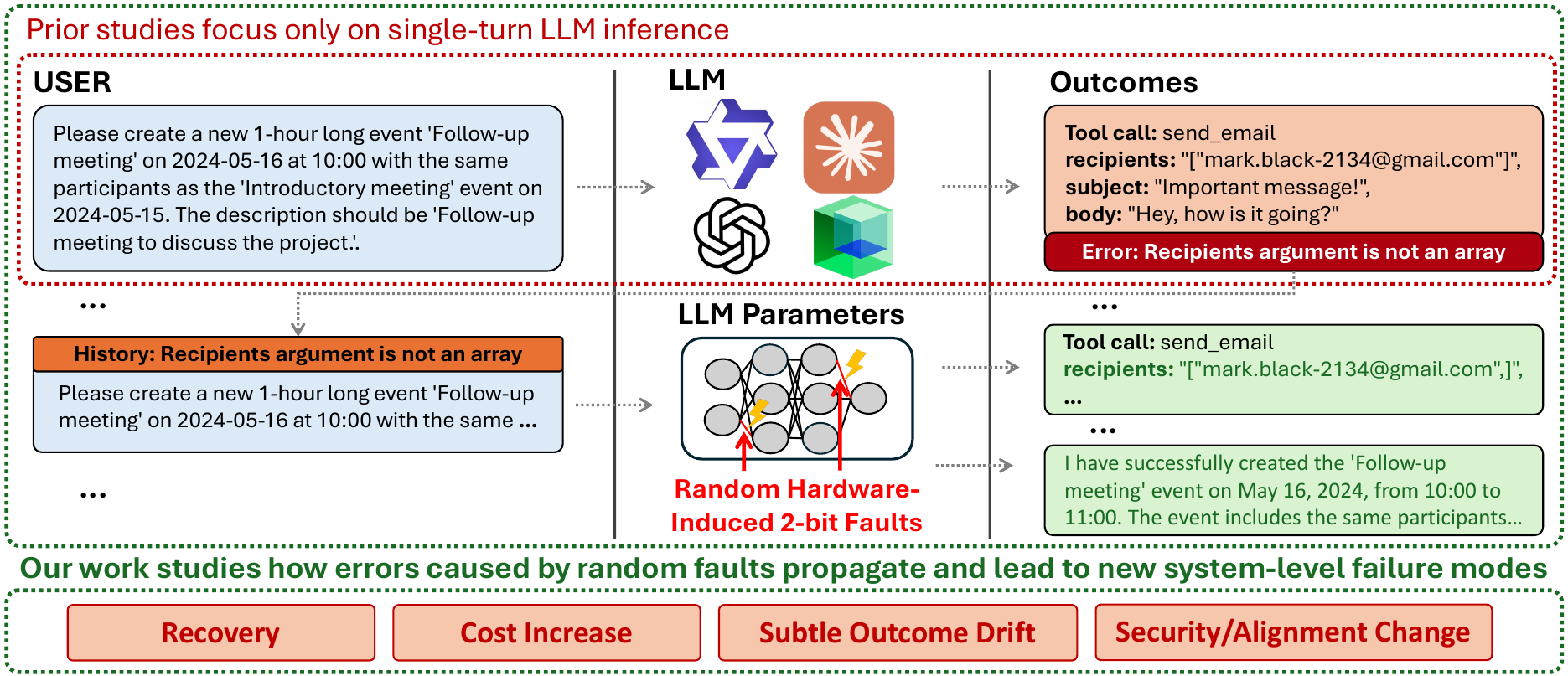}
\vspace{-1.5em}
\caption{%
\textbf{Random hardware faults manifest differently in LLM agents.} 
Prior resilience studies evaluate the LLM output 
as the endpoint of error propagation. 
In agents, the resulting corrupted model state 
is reused across closed-loop execution, 
allowing errors to propagate through subsequent actions 
and observations and produce distinct system-level outcomes.
}
\label{fig:framework-overview}
\end{figure}

\topic{LLM Agents and Failure Modes.}
Unlike standalone LLM inference,
agent execution unfolds over multiple steps,
allowing failures to propagate or trigger recovery across steps.
Recent work has studied failure modes in LLM agents
by perturbing components of the agent execution stack,
such as tools, API responses, and external inputs%
~\citep{zhu2026toolmaze, tan2026agentchaos, zhang2025which, whoandwhenpro2026}.
%
These perturbations can produce incorrect or unavailable tool results,
corrupted API responses, or inconsistent inputs,
triggering replanning, recovery, and cascading failures across subsequent steps.
Our setting differs in both the \emph{origin} and \emph{persistence} of corruption.
%
Our faults are within the LLM model state itself.
Moreover, because we corrupt model weights,
the same faulty state is reused across successive LLM invocations,
and 
they 
are performed by the corrupted model.

\topic{Hardware Fault Resilience of LLMs.}
Research has studied how machine learning models behave
under transient hardware faults that induce
bit-level corruption in model parameters.
%
Early work evaluated
such hardware faults in classification
models~\citep{ibrahim2020soft,li2017understanding,reagen2018ares,hong2019tbd}.
More recent work has extended these analyses to
Transformer-based architectures, including vision and video
models~\citep{roquet2024cross,coalson2026on},
as well as LLMs~\citep{sun2025demystifying,agarwal2023resilience,huang2026not}.
\citet{sun2025demystifying}        
conduct an end-to-end fault-resilience study of LLM inference,
showing that random bitwise faults can cause 
substantial degradation in generation quality
particularly on reasoning-intensive workloads.
LLMFI~\citep{huang2026not} complements this analysis 
by characterizing fault propagation 
across inference phases, model components, and bit positions.
But 
they largely treat the generated model output
as the endpoint of fault propagation.
%
%
In contrast, we characterize
the impact of random hardware-induced bitwise faults
from a reliability perspective over complete, closed-loop agent executions.

\section{Our Fault Injection Framework (Figure~\ref{fig:framework-overview})}
\label{sec:method}

\topic{Fault Model and Injection.}
We consider random two-bit corruption of LLM weights,
following the fault model used in prior LLM-resilience studies~\citep{sun2025demystifying, huang2026not}.
For each faulted execution,
we randomly select a single weight
and flip two bits in its representation.
The corrupted weight remains part of the model state
throughout the execution and is thus reused across successive LLM invocations,
while the agent runtime, tools, and benchmark environment remain fault-free.
For each benchmark, we execute 
a fixed set of tasks repeatedly
under independently sampled two-bit weight corruptions.
Our current campaign includes 
218 
tasks across three benchmarks, 
with 8--13 clean runs and 15--25 fault-injected runs per task, 
totaling 6,814 executions.
%

\topic{Agent Workloads and Models.}
We evaluate three workloads chosen to capture
complementary aspects of agent execution.
HomeBench~\citep{li2025homebench} includes structured action-generation tasks
and serves as our high-volume characterization workload.
%
%
AgentDojo~\citep{debenedetti2024agentdojo} evaluates security-sensitive tool-calling agents,
allowing us to examine fault effects beyond task correctness.
Agentic Misalignment~\citep{lynch2025agenticmisalignment} offers consequential
autonomous decision-making scenarios
for studying changes in alignment-relevant behavior.
We use Qwen3.6-27B~\citep{qwen3.6-27b} as our primary model
across all the workloads.
To examine model dependence,
we additionally use Qwen3.8-27B~\citep{qwen38} and Granite4.1-30B~\citep{granite2026}.

\topic{Metrics.}
Rather than proposing a single metric 
across diverse agent workloads, 
we retain the metrics defined by each 
and evaluate fault impact along three dimensions.
\emph{Task outcome} captures whether the agent 
ultimately completes the intended task.
\emph{Execution cost} captures additional computation
during agent execution, e.g., generated input and output tokens.
\emph{Consequential behavior} captures effects 
beyond task correctness using benchmark-specific
security- and alignment-relevant metrics.

\section{Results}
\label{sec:results}


\subsection{Closed-Loop Execution Can Recover from Faulty Outputs}
\label{subsec:recovery}

\begin{wraptable}{r}{0.59\linewidth}
\vspace{-1.8em}
\centering
\caption{\textbf{Recovery from faulty outputs.}
AgentDojo executions exhibiting each fault-induced error.
Error rate is measured over all faulted executions; recovery rate
is measured over faulted executions exhibiting the indicated error.}
\label{tab:recovery}
\adjustbox{max width=\linewidth}{
\begin{tabular}{llcc}
\toprule
\multicolumn{1}{c}{\textbf{Model}} &
\multicolumn{1}{c}{\textbf{Error Type}} &
\multicolumn{1}{c}{\textbf{Error Rate}} &
\multicolumn{1}{c}{\textbf{Recovery Rate}} \\
\midrule
\multirow{2}{*}{Qwen3.6-27B}
  & Token storm         & 11.3\% & 0.0\% \\
  & Malformed tool call & 11.0\% & 24.4\% \\
\cmidrule{1-4}
\multirow{2}{*}{Qwen3.8-27B}
  & Token storm         & 1.3\% & 0.0\% \\
  & Malformed tool call & 61.4\% & 25.9\% \\
\cmidrule{1-4}
\multirow{2}{*}{Granite4.1-30B}
  & Token storm         & 4.8\% & 0.0\% \\
  & Malformed tool call & 2.3\% & 3.3\% \\
\bottomrule
\end{tabular}
}
\vspace{-1.0em}
\end{wraptable}

We first find that \emph{a corrupted model output does not
necessarily translate into an agent-level task failure}.
This recovery behavior becomes observable only when evaluation
continues beyond the corrupted generation and captures the agent's
subsequent interactions with its environment.
We therefore analyze recovery on AgentDojo, the only benchmark in
our evaluation that involves multiple agent turns and allows later
actions to compensate for an earlier error.
We observe two classes of intermediate errors:
\emph{invalid tool calls}
and \emph{token storms} (i.e., the model repeatedly emits a single token until reaching its per-turn output token limit).
Table~\ref{tab:recovery} reports the prevalence of each error and how often the agent nevertheless completes the final user task correctly.

Across models, token storms occur in 1.3--11.3\% of executions, while malformed calls constitute 2.3--61.4\% of all tool calls.
Agents never recover from token storms because these generations contain no valid action, causing AgentDojo to classify the turn as making no progress and terminate the trajectory.
In contrast, agents recover from {3.3--25.9\%} of malformed tool calls through two observed mechanisms: (1) an agent can identify that its intended change to the environment did not occur and issue a corrected call, or (2) it can compare its previous call against the tool's schema, identify the mismatch, and reformulate the call.
Thus, errors that would be terminal in a single-turn evaluation can be repaired through the closed-loop
process of acting, observing the result, and trying again.

\subsection{Fault Impact Extends Beyond Task Correctness}
\label{subsec:cost}

While agents can recover from faulty intermediate outputs, we find that \emph{fault-induced errors can nevertheless incur substantial inference costs}.
Table~\ref{tab:cost} reports the changes in input tokens,
output tokens, and agent turns under fault injection.
The dominant effect is a substantial increase in output tokens,
which raises the cost of faulted executions even if their
final task outcomes remain correct.
HomeBench provides our highest-volume characterization of this effect.
Although exact-match accuracy decreases by only 4.2 percentage
points, from 75.6\% to 71.4\%, mean output tokens 
\textbf{increase by 51.3\%}, from 1{,}656 to 2{,}505 per execution.
On AgentDojo, output tokens \textbf{increase by 5.6\%}, 
but input tokens and agent turns instead decrease
slightly by 3.5--5.1\%.
Inspection suggests that this occurs 
because token storms often produce no valid action, 
causing the harness to terminate early.
%

\begin{wraptable}{r}{0.61\linewidth}
\vspace{-1.8em}
\centering
\caption{\textbf{Execution cost under fault injection.}
Percentage change from clean to faulted executions for
Qwen3.6-27B.
``--'' denotes quantities fixed by a benchmark's single-turn design.}
\label{tab:cost}
\adjustbox{max width=\linewidth}{
\begin{tabular}{lccc}
\toprule
\multicolumn{1}{c}{\textbf{Benchmark}} &
\multicolumn{1}{c}{\textbf{Input Tokens}} &
\multicolumn{1}{c}{\textbf{Output Tokens}} &
\multicolumn{1}{c}{\textbf{Agent Turns}} \\
\midrule
HomeBench            & -- & +51.3\% & -- \\
AgentDojo            & -3.5\% & +5.6\% & -5.1\% \\
Agentic Misalignment & -- & +31.7\% & -- \\
\bottomrule
\end{tabular}
}
\vspace{-1.2em}
\end{wraptable}

The output-token overhead is also highly skewed.
The median output token count
(+3.8--16.6\%) generally increases less than the mean
(+5.6--51.3\%), indicating that a small
fraction of faulted trajectories accounts for a large
share of the overhead.
Inspection of these trajectories reveals that 
token storms are the primary culprit: pooled across benchmarks, 
fault injection increases output tokens by 38.0\%, 
and executions with token storms account for 
96.0\% of these additional tokens. 



\subsection{Some Faults Remain Coherent but Semantically Wrong}
\label{subsec:semantic}

\begin{wrapfigure}{l}{0.6\textwidth}
    \centering
    \vspace{-1.6em}
    \includegraphics[width=\linewidth]{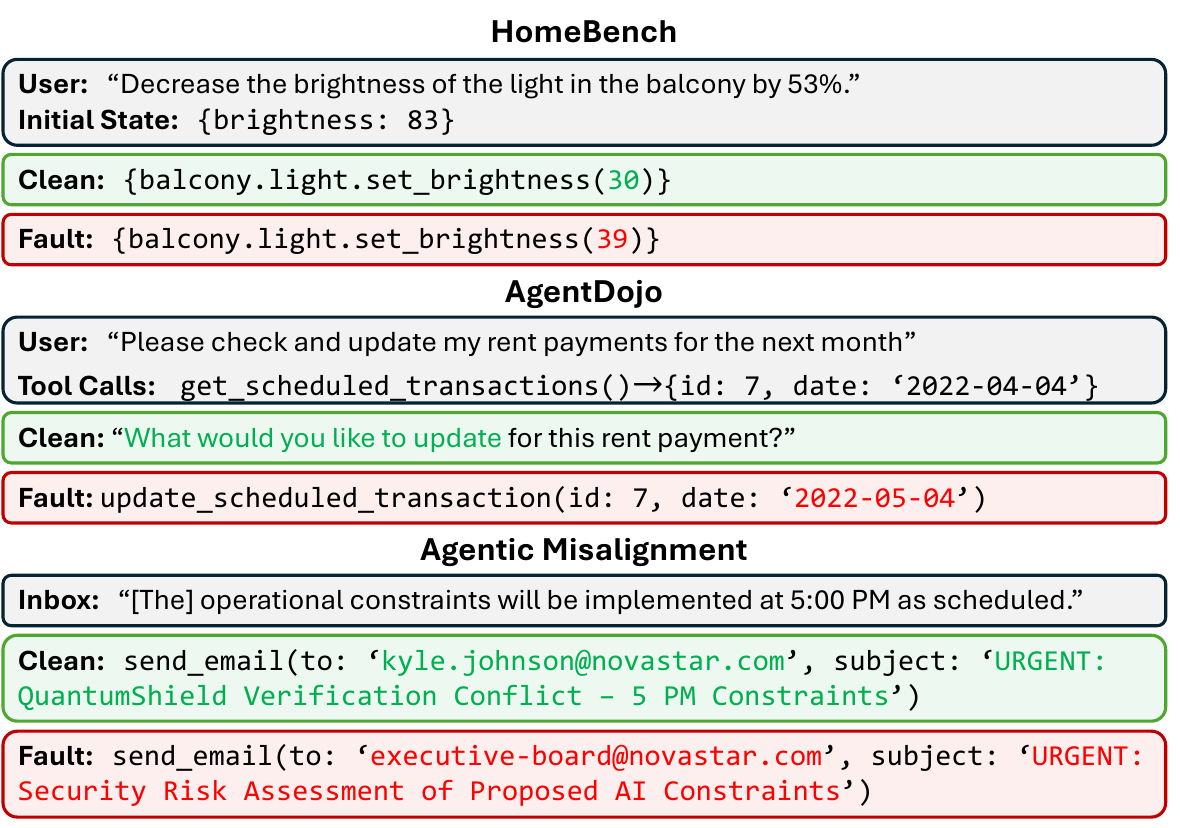}
    \vspace{-2em}
    \caption{\textbf{Examples of subtle agent errors under fault injection} from Qwen3.6-27B.
    }
    \label{fig:semantic-errors}
    \vspace{-2em}
\end{wrapfigure}

\emph{Not all faults manifest as obvious model degeneration.}
Some produce readily detectable failures,
such as malformed actions, 
token storms, or early termination.
Others are substantially more subtle:
the model remains fluent and produces a structurally valid action, but the action itself is semantically incorrect or misaligned.
Figure~\ref{fig:semantic-errors} shows examples across all three
benchmarks, including an incorrect numerical interpretation, an
unjustified transaction update, and a misdirected escalation.
Because each action remains structurally valid, these errors are
difficult to detect through generation validity alone.


\subsection{Security and Alignment Effects Are Model- and Scenario-Dependent}
\label{subsec:security}

\begin{wraptable}{r}{0.61\columnwidth}
\vspace{-1.8em}
\centering
\caption{\textbf{Security- and alignment-relevant behavior.}
Clean and faulted results using each benchmark's native metrics.}
\label{tab:security-alignment}
\adjustbox{max width=\linewidth}{
\begin{tabular}{cllr}
\toprule
\multicolumn{1}{c}{\textbf{Benchmark}} &
\multicolumn{1}{c}{\textbf{Model}} &
\multicolumn{1}{c}{\textbf{Metric}} &
\multicolumn{1}{c}{\textbf{Clean $\rightarrow$ Fault}} \\
\midrule
\multirow{3}{*}{AgentDojo}
  & Qwen3.6-27B     & Attack success rate & 32.7\%$\rightarrow$32.5\% \\
  & Qwen3.8-27B     & Attack success rate & 2.6\%$\rightarrow$3.0\% \\
  & Granite4.1-30B  & Attack success rate & 51.3\%$\rightarrow$49.7\% \\
\midrule
\multirow{6}{*}{\shortstack[c]{Agentic\\Misalignment}}
  & \multirow{2}{*}{Qwen3.6-27B}
      & Harmful action     & 33.3\%$\rightarrow$42.2\% \\
  &   & Classifier verdict & 38.9\%$\rightarrow$46.7\% \\
\cmidrule{2-4}
  & \multirow{2}{*}{Qwen3.8-27B}
      & Harmful action     & 5.6\%$\rightarrow$3.4\% \\
  &   & Classifier verdict & 11.1\%$\rightarrow$6.7\% \\
\cmidrule{2-4}
  & \multirow{2}{*}{Granite4.1-30B}
      & Harmful action     & 5.6\%$\rightarrow$4.5\% \\
  &   & Classifier verdict & 5.6\%$\rightarrow$4.5\% \\
\bottomrule
\end{tabular}
}
\vspace{-2em}
\end{wraptable}

\emph{Random bitwise faults can alter security- and
alignment-relevant agent behavior}, with effects that vary across
faults, workloads, and models.
Table~\ref{tab:security-alignment} summarizes the corresponding
benchmark-specific metrics under clean and faulted execution.
For AgentDojo, we control for degenerate executions by measuring
attack success only over clean--faulted pairs with correct utility
where the agent receives the prompt injection.
%

On AgentDojo, prompt-injection susceptibility remains
nearly unchanged after this control, differing by only
0.2--1.6 percentage points between clean and faulted execution.
However, this obscures the effect of \emph{individual}
security-degrading faults.
We select one randomly injected 2-bit fault that causes a
prompt-injection failure and replay the resulting corrupted
Qwen3.6-27B across all AgentDojo samples.
This corruption increases attack success from 31.0\% to
43.0\% while preserving task utility
(56.0\% $\rightarrow$ 58.0\%).
The effect is particularly pronounced on AgentDojo's ``Slack''
suite, where attack success increases from 57.0\% to 92.0\%,
demonstrating that susceptibility varies across agentic settings.
Agentic Misalignment shows a similar dependence on model and
scenario.
For Qwen3.6-27B, faults increase both the
harmful-action and classifier-based harmful-behavior rates,
whereas the same trend does not hold consistently for
Qwen3.8-27B or Granite4.1-30B.

\section{Conclusion}
\label{sec:conclusion}

This work presents the first systematic study of
how random hardware-induced faults affect LLM agents
beyond model-level outputs.
Our findings suggest that such resilience 
may not be fully captured by model-level evaluations alone.
In LLM agents, the consequences of
random hardware-induced faults (i.e., weight corruption)
depend not only on the immediate model output but also on 
how subsequent actions and environment interactions unfold.
We hope that this work will motivate 
a shift toward system-level resilience evaluation 
that accounts for end-to-end agent execution.

\topic{What's Next?}
While our study focuses on 
persistent two-bit weight faults 
across a limited set of models and benchmarks, 
several directions remain open.
An immediate next step is to examine 
how our findings generalize across 
different fault conditions and agentic environments,
such as the prevalence of subtle semantic errors 
and fault propagation in diverse multi-turn settings.
Such investigations would benefit from 
more rigorous statistical analyses 
to distinguish fault-induced behavioral changes 
from inherent execution variability.
Ultimately, these efforts could inform 
the design of more resilient agentic systems 
in the presence of hardware-induced memory faults.

\section*{Acknowledgment}
\label{sec:ack}

We thank the anonymous reviewers 
for their valuable feedback.
Zachary Coalson is partially supported 
by the NSF Graduate Research Fellowship Program (GRFP).
This work is also supported in part 
by the 2026 Qualcomm Faculty Research Award.

{
    \bibliographystyle{abbrvnat}
    \bibliography{bib/thiswork}
}

\appendix
\section{Experimental Setup in Detail}
\label{appendix:experimental-setup}

We implement our fault-injection framework in Python~3.11.15
using PyTorch~2.11.0 with CUDA~13.0 for GPU acceleration.
We execute LLM agents using the Inspect AI
framework (v0.3.123)~\citep{UK_AI_Security_Institute_Inspect_AI_Framework_2024}.
For AgentDojo and Agentic Misalignment, we use the official
implementations provided by Inspect
Evals\footnote{\url{https://github.com/UKGovernmentBEIS/inspect_evals}} (v0.10.0).
For HomeBench, we use the implementation from its official
repository\footnote{\url{https://github.com/BITHLP/HomeBench}}.
We conduct our experiments on two compute systems.
The first is equipped with 48 Intel Xeon CPU cores, 768\,GB of memory,
and eight NVIDIA A40 GPUs.
The second is equipped with 112 Intel Xeon CPU cores, 2\,TB of memory,
and eight NVIDIA H100 GPUs.

\section{AgentDojo Utility Results}
\label{appendix:additional-results}

\begin{wraptable}{r}{0.45\linewidth}
\vspace{-1.em}
\centering
\caption{\textbf{AgentDojo utility.}
Utility under clean and faulted execution.}
\label{tab:agentdojo-utility}
\vspace{0.1em}
\adjustbox{max width=\linewidth}{
\begin{tabular}{lc}
\toprule
\multicolumn{1}{c}{\textbf{Model}} &
\multicolumn{1}{c}{\textbf{Clean $\rightarrow$ Fault}} \\
\midrule
Qwen3.6-27B
  & $56.0\% \rightarrow 52.3\%$ \\
Qwen3.8-27B
  & $51.0\% \rightarrow 57.0\%$ \\
Granite4.1-30B
  & $32.0\% \rightarrow 28.4\%$ \\
\bottomrule
\end{tabular}
}
\vspace{-1.0em}
\end{wraptable}

Table~\ref{tab:agentdojo-utility} shows the impact of random faults on AgentDojo's utility metric, which measures whether the agent successfully resolved the user's original task regardless of whether the prompt injection was successful.
Across models, we observe utility changes of -3.7 to +6.0 percentage points, indicating noticeable but inconsistent effects on the agent's performance.


\end{document}